\documentclass[12pt]{article}
\usepackage{graphicx}
\usepackage{amsmath}

\title{ Rare Muon Decay $\mu^+ \to e^+e^-e^+\nu_e\bar{\nu_\mu}$}

\author{\textbf{Rashid M. Djilkibaev$^{1}$ \thanks{Permanent address:
     Institute for Nuclear Research, 60-th Oct. pr. 7a,
     Moscow 117312, Russia}\ ,
     Rostislav V. Konoplich$^{1,2}$}\\
   \normalsize$^{1}$Department of Physics, New York University,
   New York, NY 10003\\
   \normalsize$^{2}$Manhattan College, Riverdale, New York, NY, 10471}

\begin{document}

\maketitle

\begin{abstract}
An analytical expression for the spin-averaged amplitude squared of
the rare muon decay $\mu^+ \to e^+e^-e^+\nu_e\bar{\nu_\mu}$ is calculated.
Monte Carlo phase space simulation using the analytical expression 
for the amplitude
has been used 
to get various differential distributions of charged leptons.
The approximate analytical expression for the total energy spectrum 
of charged leptons near the end point is presented. 
The dependence of branching ratio on cuts in total energy of charged leptons 
is studied taking into account an experimental energy resolution. It is
shown that the measured branching ratio is very sensitive to the energy resolution.

\end{abstract}

\newpage

\section*{} 

The observation of lepton flavor violation (LFV) processes  would indicate  new 
physics beyond the Standard Model \cite{ver}, \cite{kuno}.
The lepton flavor violation effects in processes of
$\mu \to e$ conversion in a muonic atom, radiative muon decay
$\mu^+ \to e^+\gamma$ and muon decay $\mu^+ \to e^+e^-e^+$
may be large enough to be detected in the future experiments.
Ideas \cite{new} of improvement of muon beam intensity by a few orders 
of magnitude lead to new possibilities  
in designing of new lepton flavor violation experiments.
Analysis of rare background processes is very important for understanding the
feasibility of such experiments.

\noindent
In this article we consider the  process

\begin{equation}
\mu^+ \to e^+e^-e^+\nu_e\bar{\nu_\mu}
\label{decay} 
\end{equation}

which is the most important background process in searches for
the lepton flavor violating muon decay $\mu^+ \to e^+e^-e^+$.
The current measured branching ratio for process (\ref{decay})
is $\Gamma(\mu^+ \to e^+e^-e^+\nu_e\bar{\nu_\mu})/\Gamma_{total} = (3.4 \pm 0.4) \times 10^{-5}$ ~~\cite{pdg},\cite{bertl}.

The detailed analysis of background processes for
new experiments 
searching for the muon decay $\mu^+ \to e^+e^-e^+$
requires  knowledge of the amplitude of  process (\ref{decay}).  

The branching ratio of process (\ref{decay}) was first calculated  in \cite{jinrprep} and \cite{bardin}
and later in \cite{saper}, \cite{gaem} and \cite{kersch}. Unfortunately works \cite{jinrprep} and
\cite{saper} were not published, and articles \cite{bardin} and  \cite{gaem}
give only final results of calculations for specific experimental 
cuts or without cuts. In \cite{kersch} the analysis of  process (\ref{decay})
was performed but the final expression for the
squared amplitude was not given.

\begin{figure}[htb!]
  \centering{\hbox{
  \includegraphics[width=0.5\textwidth]{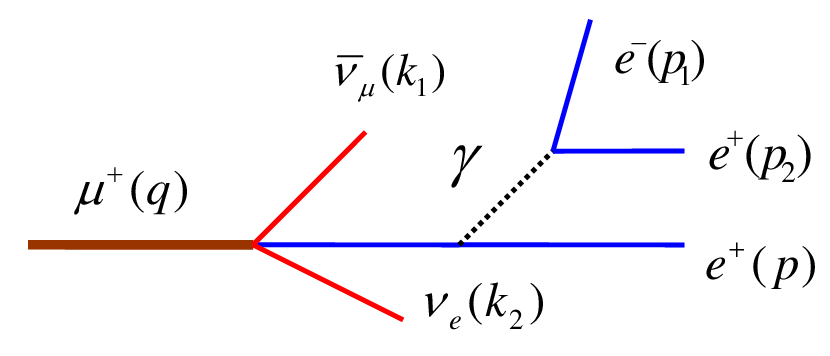}
  \includegraphics[width=0.5\textwidth]{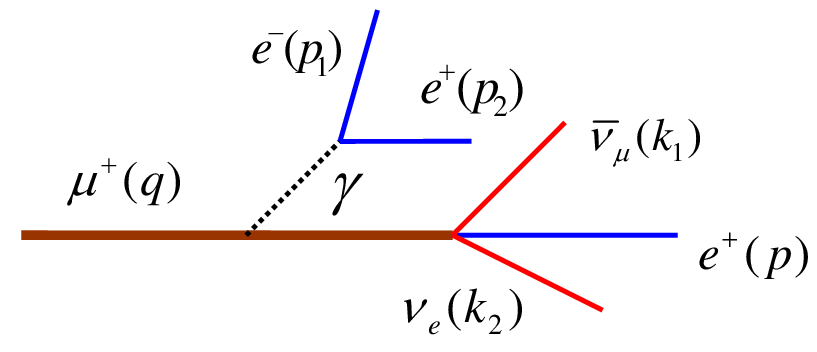} }}
  \caption[Short caption.]{Two Feynman diagrams for the process (\ref{decay}).
The other two diagrams are obtained by the interchange $p \leftrightarrow p_2$.}
\label{feyndiag}
\end{figure}

In order to allow calculation of the
branching ratio of process (\ref{decay}) for any desired cuts and parameters 
an analytical expression for spin-averaged amplitude squared 
$\overline{|\tilde{F}|^2}$ 
is presented in the Appendix.
The amplitude was calculated for the case of a V-A interaction and unpolarized muons
by using Mathematica 6.0 and FeynCalc 6.0.0 \cite{feyncalc}.
The branching ratio calculation of process (\ref{decay})  combines
$\overline{|\tilde{F}|^2}$ with a phase space Monte Carlo simulation.
In comparison 
with the previous articles (such as \cite{gaem} using the $1.3 \times 10^6$ events) 
the statistics was significantly increased
to $10^8$ and $5 \times 10^8$ events 
in dependence on cuts applied to
charged lepton energies. 

In the lowest order in perturbation theory process (\ref{decay})
is described by four Feynman diagrams, two of which are shown in 
Fig.\ref{feyndiag}.
The other two are obtained by the interchange $p \leftrightarrow p_2$.
Additional diagrams corresponding to  particle emission from internal
W-boson lines are suppressed by a factor $\sim (m_{\mu}/M_W)^2 \sim 10^{-6}$.

Below we present the results of these calculations of the branching ratio 

\begin{equation}
R \equiv \frac{\Gamma(\mu^+ \to e^+e^-e^+\nu_e\bar{\nu_\mu})}{\Gamma(\mu^+ \to e^+\nu_e\bar{\nu_\mu})}
\label{rate} 
\end{equation}
   
\begin{figure}[htb!]
  \centering{\hbox{
  \includegraphics[width=0.47\textwidth]{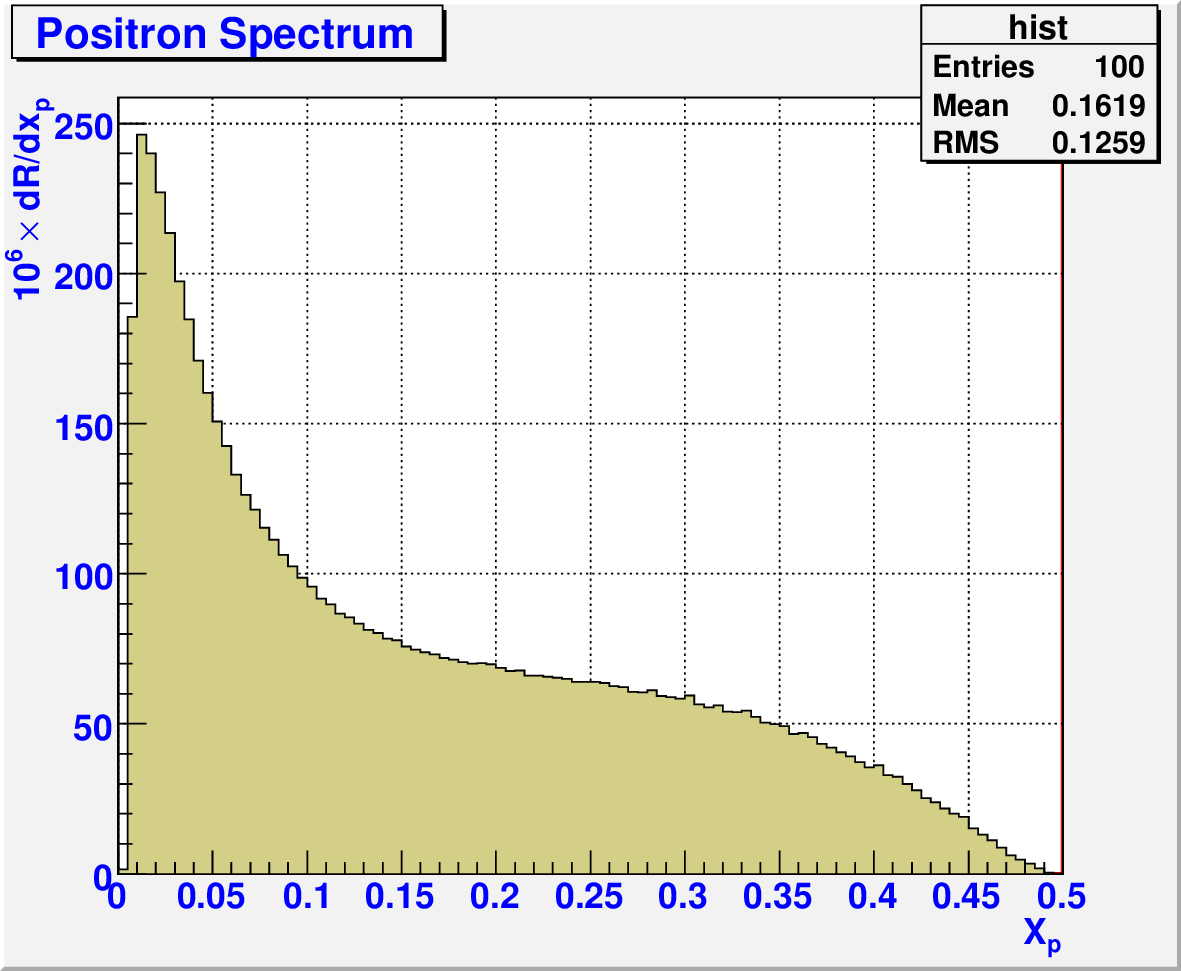}
  \includegraphics[width=0.47\textwidth]{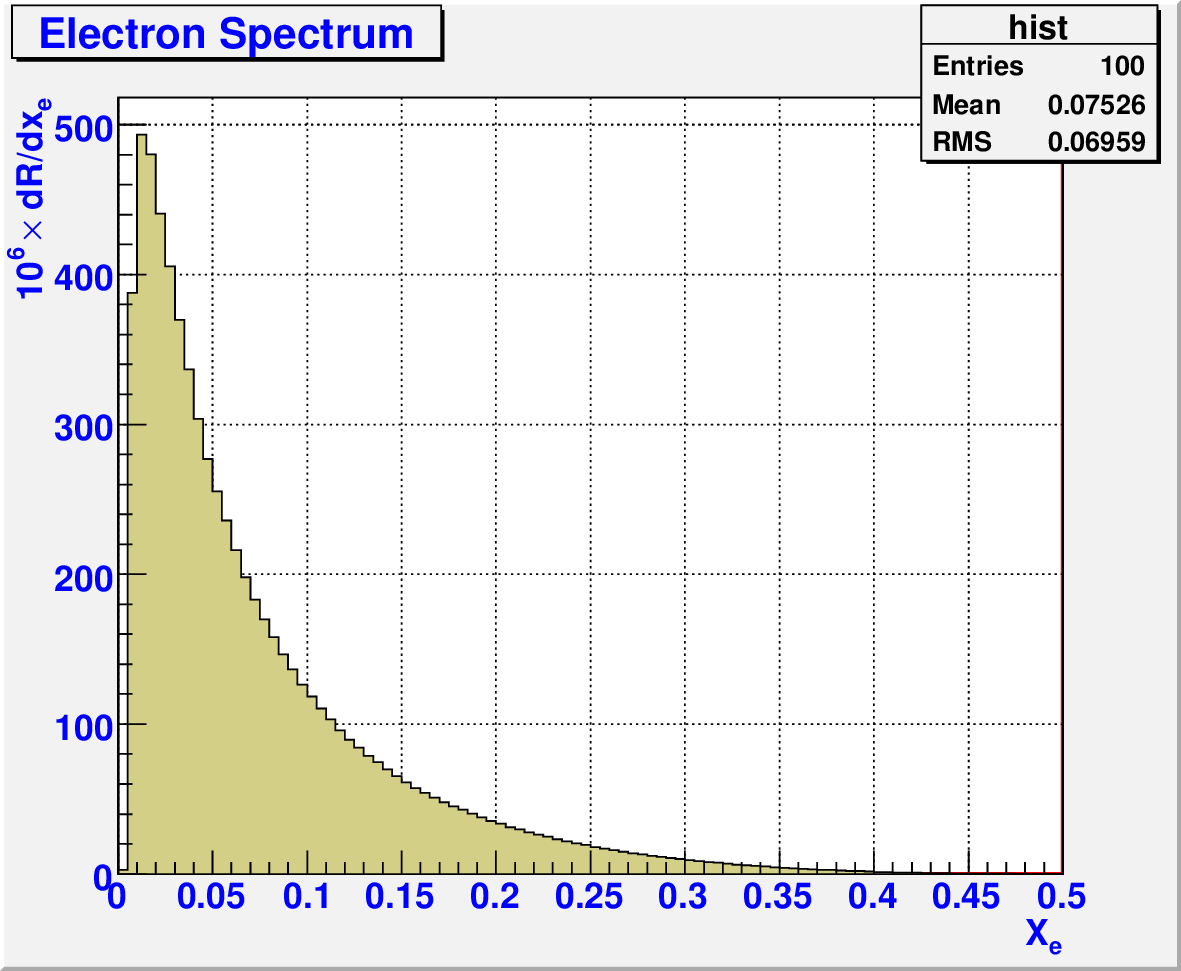} }}
  \caption[Short caption.]{The distribution in positron energy $dR/dx_p$ (left)
and the distribution in electron energy $dR/dx_e$ (right) where $x_p = E_p/m_{\mu}$ and $x_e = E_e/m_{\mu}$}
\label{ep_50m}
\end{figure}

\begin{figure}[htb!]
  \centering{\hbox{
  \includegraphics[width=0.47\textwidth]{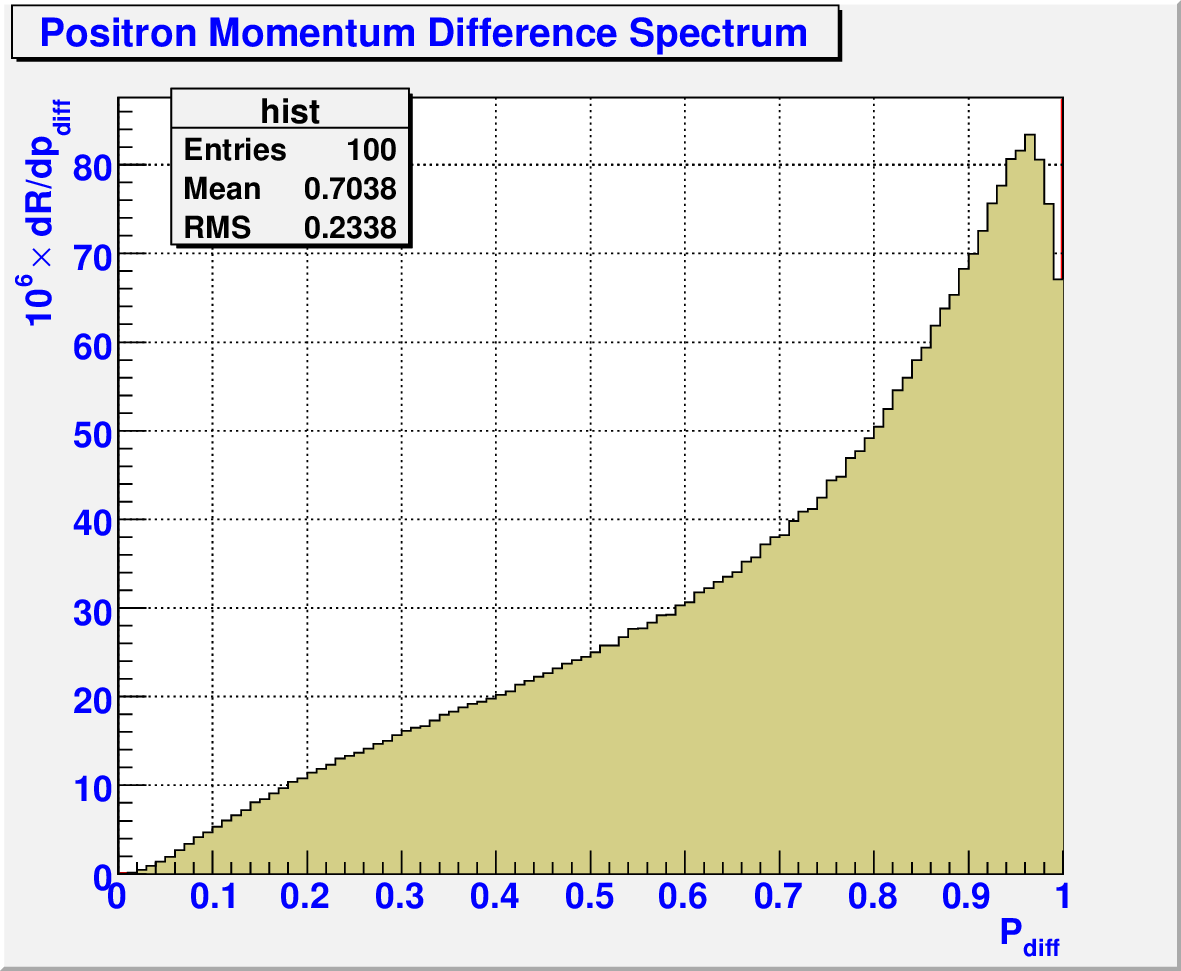}
  \includegraphics[width=0.47\textwidth]{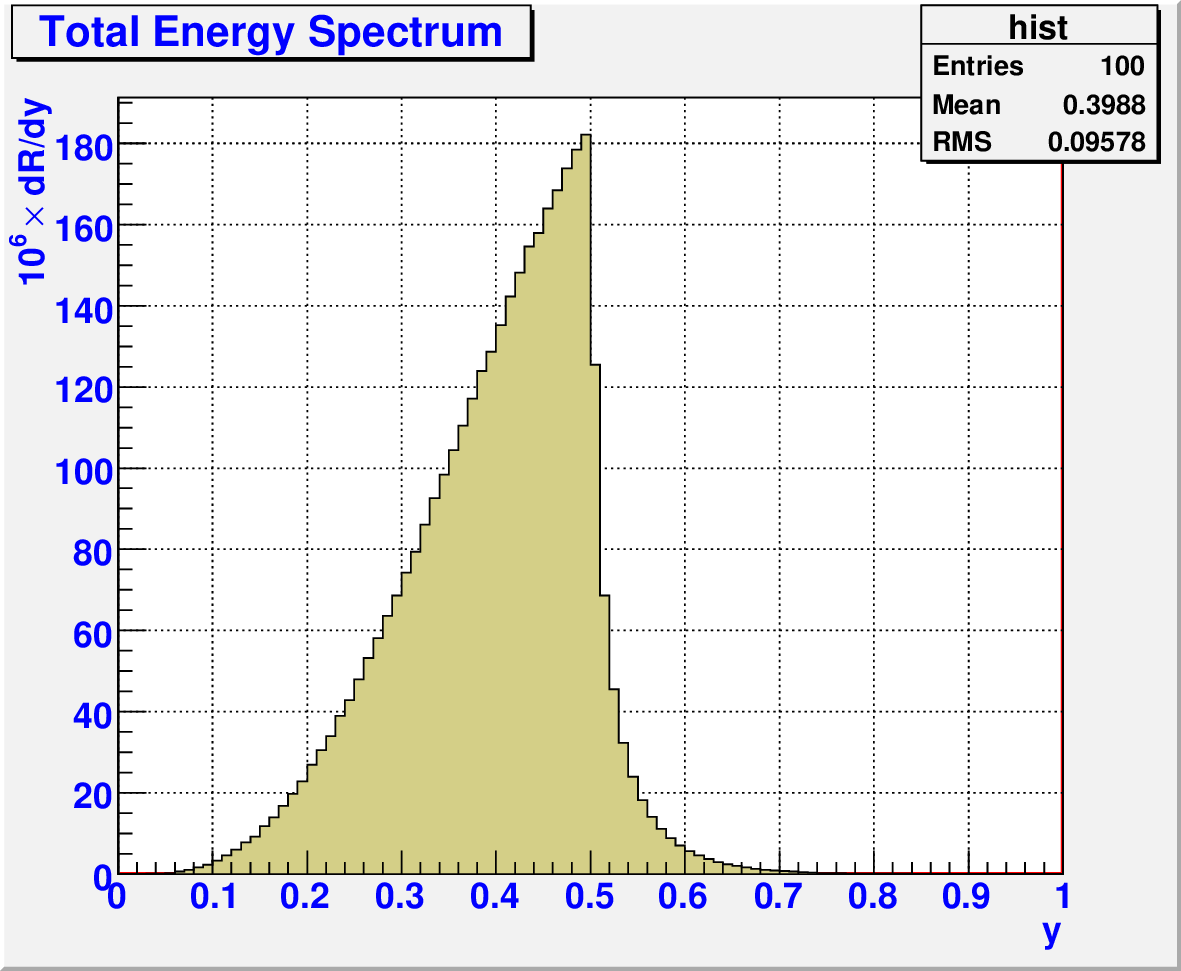} }}
  \caption[Short caption.]{The distribution in positron variable $dR/dp_{diff}$ (left)
and the distribution in total energy of charged leptons (right)  where $y = E_{tot}/m_{\mu}$.
}
\label{diff_50m}
\end{figure} 

Distributions in positron energy and 
electron energy without cuts for process (\ref{decay}) 
are shown in Fig.\ref{ep_50m}. The positron 
distribution is characterized
by a long tail up to the maximum positron energy. The electron distribution
peaks near zero and decreases quickly as electron energy tends to its
upper limit. 

The distribution in relative positron momentum difference 
$p_{diff} \equiv |\overrightarrow{p} - \overrightarrow{p_2}|/
|\overrightarrow{p} + \overrightarrow{p_2}|$ where 
$p$ and $p_2$ are positron momenta is shown in Fig.\ref{diff_50m} (left). 
The $p_{diff}$ distribution vanishes at $p_{diff} = 0$ in agreement 
with Fermi statistics of positrons.
%Zero value of distribution
%in $p_{diff}$ is in agreement with Fermi statistics of positrons.
A distribution in total energy $E_{tot}$ of all 
charged leptons without cuts is shown in Fig.\ref{diff_50m} (right). 
The spectrum is steep in the region above $0.5m_{\mu}$.
This is important for background suppression in searches for
lepton flavor violating process $\mu^+ \to e^+e^-e^+$.

\begin{table} [htb!]
\begin{center}
 \begin{tabular}{|c|c|c|}
   \hline
   Cut $(m_{\mu} - E_{tot})$ & Statistics & Branching ratio R  \\
   \hline\hline
   $1m_e$ & $5 \times 10^8$ & $(2.83 \pm 0.16)\times 10^{-19}$ \\
   $5m_e$ & $5 \times 10^8$ & $(4.660 \pm 0.046)\times 10^{-15}$ \\
   $10m_e$ & $10^8$ & $(3.091 \pm 0.032) \times 10^{-13}$ \\
   $50m_e$ & $10^8$ & $(7.127 \pm 0.013)\times 10^{-9}$ \\
   $100m_e$ & $10^8$ & $(2.1123  \pm 0.0022)\times 10^{-6}$ \\
   no cut & $10^8$ & $(3.5908 \pm 0.0033)\times 10^{-5}$ \\
   \hline
 \end{tabular}
\caption{The branching ratio R versus cut
$(m_{\mu} - E_{tot})$ on the total energy of charged leptons.
The cut is measured in the electron mass $m_e$.
}\label{tab:results}
\end{center}
\end{table} 

\begin{figure}[htb!]
  \centering{\hbox{
  \includegraphics[width=0.5\textwidth]{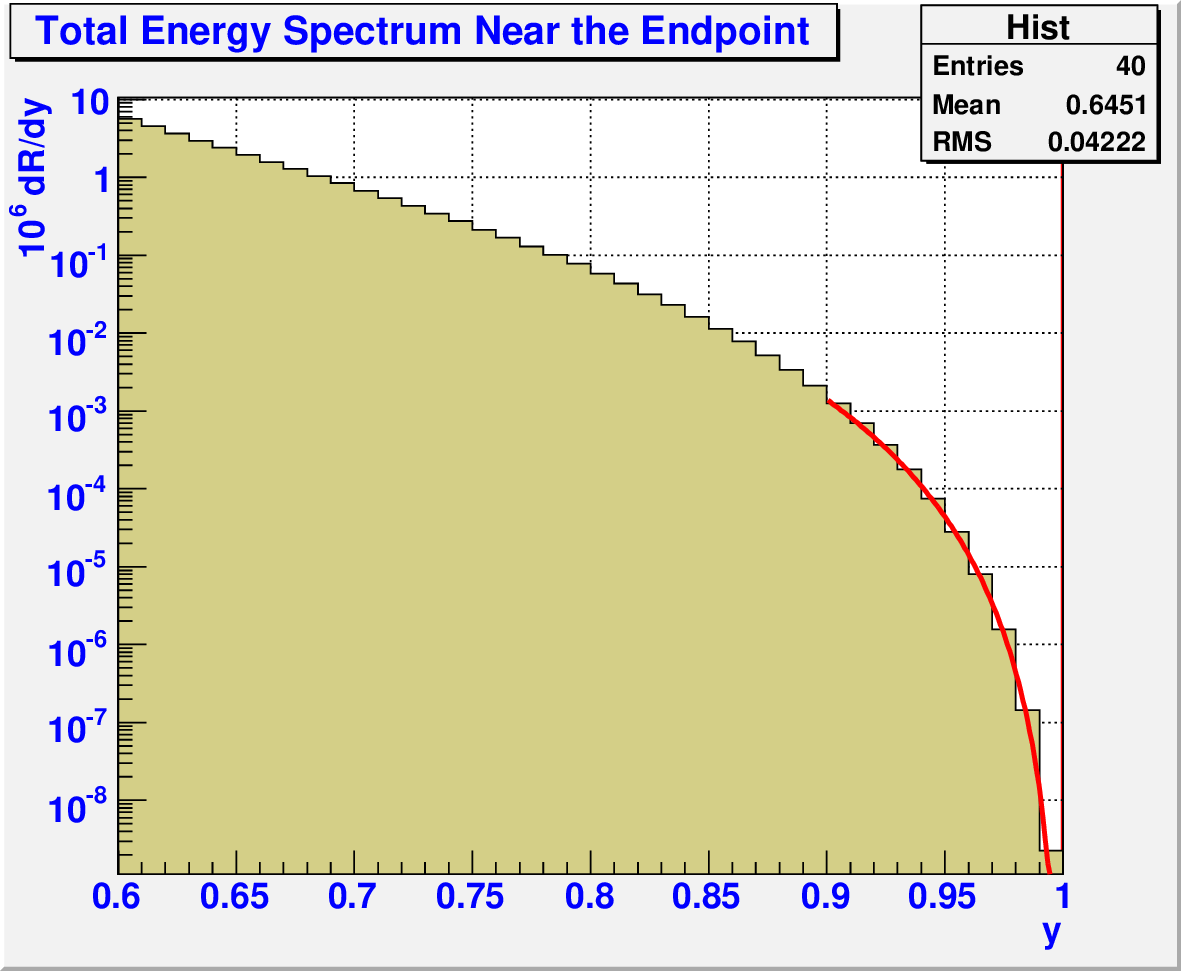} 
  \includegraphics[width=0.5\textwidth]{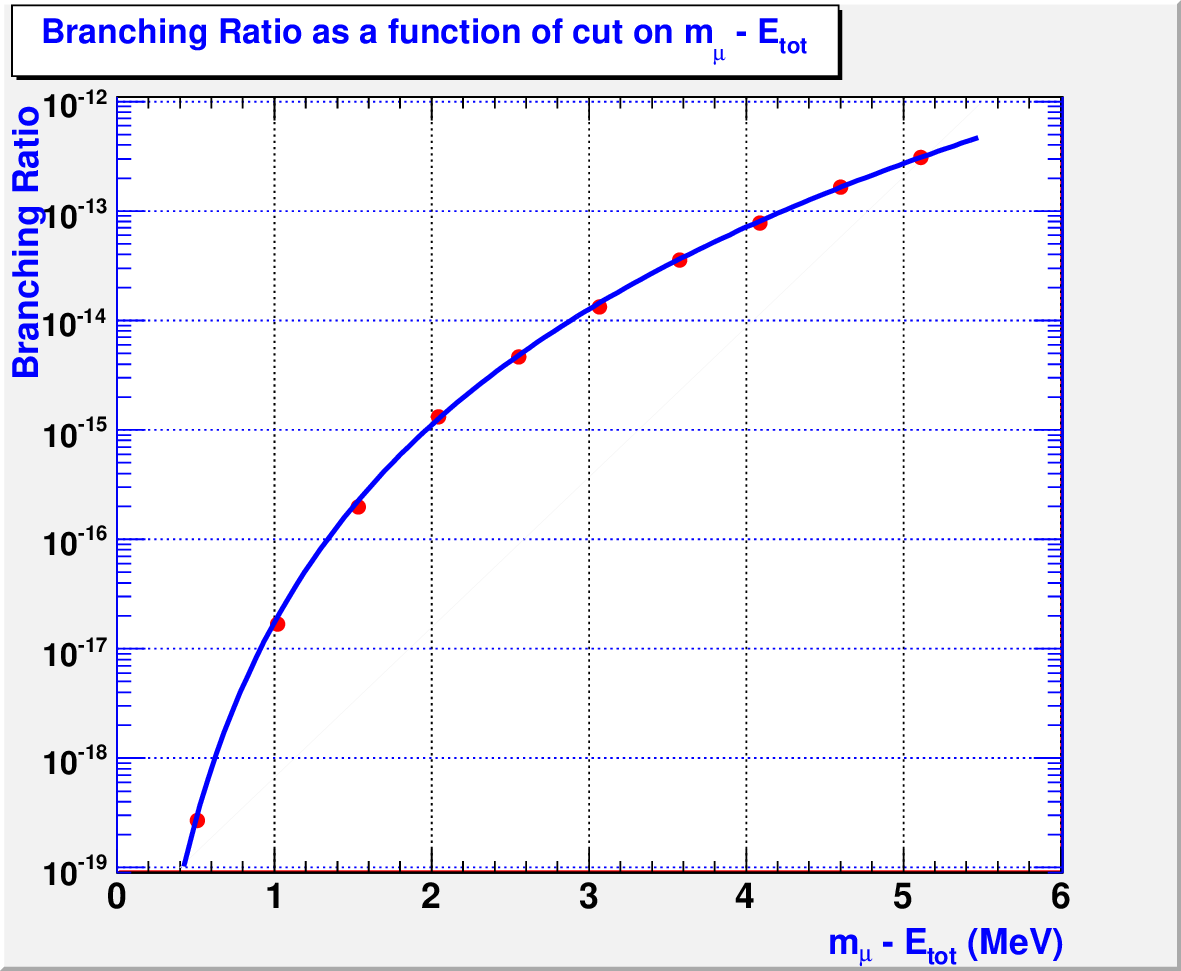} }}
  \caption[Short caption.]{
The total energy spectrum dR/dy near the endpoint (left) and the  
branching ratio R versus  cut
on the total energy of charged leptons (right) where $y = E_{tot}/m_{\mu}$. 
Smooth lines show the results of fit.}
\label{results}
\end{figure}

\noindent
The results of calculations of the branching ratio 
R are presented 
in Table \ref{tab:results} and Fig.\ref{results} (right)
versus cut on $(m_{\mu} - E_{tot})$
where $E_{tot}$ is the total energy of all charged leptons.
Without cuts our result is in agreement with the results of articles 
\cite{bardin}, \cite{gaem} and \cite{kersch}:
$R = (3.54 \pm 0.09) \times 10^{-5}$, $R = (3.5916 \pm 0.0022) \times 10^{-5}$ 
and $R = (3.6 \pm 0.1) \times 10^{-5}$ respectively. For the cut $(m_{\mu} - E_{tot}) = 5m_e$
our result is close to those of \cite{gaem} but for cuts of 10, 50, 100$m_e$
our results are less by a factor 10. 
We note that the differential distribution $dR/dE_{tot}$ of \cite{gaem}
contradicts the results quoted in their table.

%\vspace{-0.3in}
 
Also, we would like to note that uncertainties in \cite{gaem} appear
quite small for the given number of trials. Our statistics is 
about 100 times higher but uncertainties are comparable with \cite{gaem}. 

%\noindent
Note that the cut $< 5m_e$ (see first lines in  Table \ref{tab:results})
corresponds to the region of interest in searches for the lepton flavor
violating process $\mu^+ \to e^+e^-e^+$.
By applying fit to the region near the endpoint
(see Fig.\ref{results}(left)) an approximate expression for
the spectrum in this region was found to be

\begin{equation}
dR/dy  = C \cdot  (1 - y)^5 
\label{fit} 
\end{equation}
where $C = 1.4 \times 10^{-4}$ and $y = E_{tot}/m_{\mu}$.

Fit of the branching ratio R (see Fig.\ref{results}(right)) near the endpoint 
gives an approximate expression
\begin{equation}
R = 2.99 \cdot 10^{-19} ~ \big (\frac{m_{\mu} - E_{tot}}{m_e} \big)^6
\label{fit1} 
\end{equation}
Below we show that the steep behavior of the total energy
spectrum near the endpoint with a finite detector resolution
lead to a significant increase in the number
of background events of type (\ref{decay}) in searches for lepton flavor violating muon decay $\mu^+ \to e^+e^-e^+$.

The branching ratio  to have a signal from muon decay (\ref{decay})  above a threshold
$m_{\mu} - \Delta$ is given by

\begin{equation}
R_{exp} = \int\limits_{m_{\mu} - \Delta}^\infty dE^{m}_{tot} \int\limits_0^{m_{\mu}} dR/dE_{tot} \cdot f(E^{m}_{tot} - E_{tot}) dE_{tot}
\label{eqP}
\end{equation}

where $E^{m}_{tot}$ is the measured total energy of charged particles, $E_{tot}$ is the true total energy, 
$\Delta$ is a value  measured from the total energy endpoint,
$dR/dE_{tot}$ is the total energy spectrum near the endpoint, and f is the resolution function of the detector.

For the detector response function of the Gaussian form
$f(x) = \frac{1}{\sqrt{2\pi}\cdot \sigma} ~~exp(~ -\frac{x^2}{2\sigma^2}~) $
the branching ratio $R_{exp}$,
given by Eq.(\ref{eqP}), can
be calculated analytically by using expression (\ref{fit})
for the total energy spectrum. In this case

\begin{equation}
R_{exp} =  \frac{C}{6\sqrt{2 \pi}} ~(\frac{\sigma}{m_{\mu}})^6~ I .
\label{eqPerf}
\end{equation}

The factor I is given by

\begin{equation}
I = \sqrt{\frac{\pi}{2}}(15 + 45u^2 + 15u^4 + u^6)Erfc(\frac{-u}{\sqrt{2}}) +
u(33 + 14u^2 + u^4)e^{-u^2/2} ,
\end{equation}

where $u = \Delta/\sigma$, $Erfc(z) = 1 - \frac{2}{\sqrt{\pi}}\int\limits_0^z e^{-t^2} dt $

In the limit  $\Delta >> \sigma$  (an ideal detector) the branching ratio is 
given by $R_{exp} = \frac{C}{6} (\frac{\Delta}{m_{\mu}})^6$.
In the case $\Delta \simeq \sigma$ the branching ratio is very sensitive to the resolution $\sigma$ and
it is proportional to $\sigma^6$. 
Due to the steep total energy spectrum near the endpoint the
detector resolution changes significantly the measured branching ratio $R_{exp}$.
For example for the typical experimental $2\sigma$ cut on the  total energy  $|m_{\mu} - E_{tot}| < 2\sigma$, 
with $\sigma = \Delta/2$ and $\Delta$ = 5$m_e$, the branching ratio $R_{exp}$ is increased by a factor 7.8 
from $4.66 \times 10^{-15}$ (see Table \ref{tab:results}) to $3.64 \times 10^{-14}$.

\section*{Acknowledgments}  
The authors thank A.Mincer and P.Nemethy
for reading the manuscript and useful suggestions.

\newpage

\section*{Appendix. The branching ratio for Muon Decay 
$\mu^+ \to e^+e^-e^+\nu_e\bar{\nu_\mu}$.}

The kinematical variables of the process (\ref{decay}) were generated by using
PhaseSpace.C program of CERN ROOT software package \cite{root} and a weight equal to the
phase space factor multiplied by the squared amplitude was assign to each event.
In this program the phase space volume is defined as 

\begin{equation}
d\Phi = \delta^{(4)}(P_f - P_i)\prod_{k=1}^N {(\frac{d^3p_k}{2p_k})}^N
\end{equation}

where $P_f$ and $P_i$ are 4-momenta of final and initial states of the 
system, N is the number of particles in the final state.

For the process (\ref{decay}) the branching ratio R of the process can be
presented in the form

\begin{equation}
R = C \times \overline{|\tilde{F}|^2}~d\Phi
\end{equation} 

where $\overline{|\tilde{F}|^2}$ is the normalized spin-averaged amplitude
of the process squared.

The following parameters were taken from
Particle Data Group tables \cite{pdg}:
muon mass $u = 0.105658367 ~GeV$; electron mass $m = 0.000510998910 ~GeV$;
$\alpha = 1/137.035999679$.

In terms of these parameters constant $C$ from the previous equation
is given by
\begin{equation}
C = 384\alpha^2/(\pi^6\mu^6).
\end{equation} 

Also we define $u2 \equiv u^2$, $m2 \equiv m^2$, $m4 \equiv m^4$, 
$titj \equiv t_it_j$ which is the scalar product of 
4-momenta t($q, p, p_1, p_2, k_1, k_2$) as shown in Fig.\ref{feyndiag}. 

qps = qp*qp;
      
qp12 = qp1*qp1;
      
qp22 = qp2*qp2;
      
pp12 = pp1*pp1;
      
pp22 = pp2*pp2;
      
p1p22 = p1p2*p1p2;

C1 = 1.0/(2.0*(m2 + pp1 + pp2 + p1p2));
      
C2 = 1.0/(2.0*(m2 - qp1 - qp2 + p1p2));
      
C3 = 1.0/(2.0*(m2 - qp - qp1 + pp1));
      
D1 = 1.0/(2.0*(m2 + p1p2));
      
D2 = 1.0/(2.0*(m2 + pp1));

tr11 = -(qk2*(p2k1*(pp12 - pp1*(m2 + pp2) + m2*(m2 + p1p2) -
	         pp2*(2.*m2 + p1p2)) + p1k1*(m4 - m2*pp2 + pp22 + m2*p1p2 -
             pp1*(2.*m2 + pp2 + p1p2)) + pk1*((2.*m2 - pp2)*(m2 + p1p2) -
             pp1*(m2 + 2.*pp2 + p1p2))));

tr12 = m2*pk1*p1k2*qp - m2*p1k1*p1k2*qp + m2*pk1*p2k2*qp -
             m2*p2k1*p2k2*qp - 2.*m2*pk1*qk2*qp - m2*p1k1*qk2*qp -
             m2*p2k1*qk2*qp + pk1*p1k2*qp*p1p2 + p2k1*p1k2*qp*p1p2 +
             pk1*p2k2*qp*p1p2 + p1k1*p2k2*qp*p1p2 - 2.*pk1*qk2*qp*p1p2 -
             p1k1*qk2*qp*p1p2 - p2k1*qk2*qp*p1p2 + qk1*(m2*qk2*pp1 +
             m2*p2k2*pp2 + m2*qk2*pp2 - p2k2*pp1*p1p2 + qk2*pp1*p1p2 +
             qk2*pp2*p1p2 - 2.*m2*pk2*(m2 + p1p2) + p1k2*(m2*pp1 - pp2*p1p2)) -
             m2*pk1*pk2*qp1 + m2*p1k1*pk2*qp1 + pk1*p2k2*pp1*qp1 +
             2.*p2k1*p2k2*pp1*qp1 - p2k1*qk2*pp1*qp1 - pk1*p2k2*pp2*qp1 -
             2.*p1k1*p2k2*pp2*qp1 + 2.*pk1*qk2*pp2*qp1 + p1k1*qk2*pp2*qp1 -
             pk1*pk2*p1p2*qp1 - p2k1*pk2*p1p2*qp1 - m2*pk1*pk2*qp2 +
             m2*p2k1*pk2*qp2 - pk1*p1k2*pp1*qp2 - 2.*p2k1*p1k2*pp1*qp2 +
             2.*pk1*qk2*pp1*qp2 + p2k1*qk2*pp1*qp2 + pk1*p1k2*pp2*qp2 +
             2.*p1k1*p1k2*pp2*qp2 - p1k1*qk2*pp2*qp2 - pk1*pk2*p1p2*qp2 -
             p1k1*pk2*p1p2*qp2 + k1k2*(2.*m2*qp*(m2 + p1p2) + pp2*(p1p2*qp1 -
             m2*qp2) + pp1*(-(m2*qp1) + p1p2*qp2));

tr13 = 2.*qk2*(p1k1*pp2*(-2.*m2 + pp2) + pk1*(pp1*(m2 - pp2) +
             m2*(m2 + p1p2) - pp2*(2.*m2 + p1p2)) + p2k1*(pp1*(m2 - pp2) +
             m2*(m2 + p1p2) - pp2*(2.*m2 + p1p2)));

tr14 = (m2*pk1*p1k2*qp + m2*p1k1*p1k2*qp + 4.*m2*p2k1*p1k2*qp -
             m2*pk1*p2k2*qp - m2*p1k1*p2k2*qp - 2.*m2*pk1*qk2*qp -
             2.*m2*p1k1*qk2*qp - 4.*m2*p2k1*qk2*qp - 2.*p1k1*p1k2*pp2*qp +
             2.*p1k1*qk2*pp2*qp + 2.*pk1*p1k2*qp*p1p2 + 2.*p2k1*p1k2*qp*p1p2 -
             2.*pk1*qk2*qp*p1p2 - 2.*p2k1*qk2*qp*p1p2 -
             qk1*(-2.*(m2 + pp1)*(m2*p2k2 - qk2*pp2) - p1k2*(pp1*(m2 + 2.*pp2) +
             m2*(m2 + pp2 - p1p2)) + m2*pk2*(m2 + pp1 + pp2 + p1p2)) -
             m2*pk1*pk2*qp1 - m2*p1k1*pk2*qp1 - 4.*m2*p2k1*pk2*qp1 +
             m2*pk1*p2k2*qp1 - m2*p1k1*p2k2*qp1 + 2.*m2*p2k1*p2k2*qp1 +
             2.*m2*pk1*qk2*qp1 + 2.*m2*p1k1*qk2*qp1 + 4.*m2*p2k1*qk2*qp1 +
             2.*pk1*p2k2*pp1*qp1 + 2.*p2k1*p2k2*pp1*qp1 + 2.*p1k1*pk2*pp2*qp1 -
             2.*p2k1*qk2*pp2*qp1 - 2.*pk1*pk2*p1p2*qp1 - 2.*p2k1*pk2*p1p2*qp1 +
             m2*pk1*pk2*qp2 + m2*p1k1*pk2*qp2 - m2*pk1*p1k2*qp2 +
             m2*p1k1*p1k2*qp2 - 2.*m2*p2k1*p1k2*qp2 + 2.*m2*pk1*qk2*qp2 +
             2.*m2*p2k1*qk2*qp2 - 2.*pk1*p1k2*pp1*qp2 - 2.*p2k1*p1k2*pp1*qp2 +
             2.*pk1*qk2*pp1*qp2 + 2.*p2k1*qk2*pp1*qp2 +
             k1k2*(m2*qp*(m2 + pp1 + pp2 + p1p2) - (pp1*(m2 + 2.*pp2) +
             m2*(m2 + pp2 - p1p2))*qp1 - 2.*m2*(m2 + pp1)*qp2))/2.0;

tr22 = -(pk1*(-(p1k2*(m2*u2 + p1p2*(u2 + qp1) + qp1*(2.*m2 - qp2) +
             m2*qp2 + qp22)) + qk2*(qp1*(m2 - 2.*qp2) + m2*(m2 + u2 + qp2) +
             p1p2*(m2 + u2 + qp1 + qp2)) - p2k2*(qp12 + qp1*(m2 - qp2) +
             p1p2*(u2 + qp2) + m2*(u2 + 2.*qp2))));

tr23 = (-2.*m2*pk1*p1k2*qp + m2*p1k1*p1k2*qp - m2*p2k1*p1k2*qp +
             m2*p1k1*p2k2*qp + m2*p2k1*p2k2*qp + 2.*m2*pk1*qk2*qp +
             2.*m2*p2k1*qk2*qp - 2.*pk1*p1k2*qp*p1p2 - 2.*p2k1*p1k2*qp*p1p2 +
             2.*pk1*qk2*qp*p1p2 + 2.*p2k1*qk2*qp*p1p2 -
             qk1*(-2.*(m2*pk2 - qk2*pp2)*(m2 + p1p2) + m2*p2k2*(m2 + pp1 +
             pp2 + p1p2) - p1k2*(m2*(m2 - pp1 + pp2) + (m2 + 2.*pp2)*p1p2)) +
             2.*m2*pk1*pk2*qp1 - m2*p1k1*pk2*qp1 + m2*p2k1*pk2*qp1 -
             4.*m2*pk1*p2k2*qp1 - m2*p1k1*p2k2*qp1 - m2*p2k1*p2k2*qp1 +
             4.*m2*pk1*qk2*qp1 + 2.*m2*p1k1*qk2*qp1 + 2.*m2*p2k1*qk2*qp1 -
             2.*pk1*p2k2*pp1*qp1 - 2.*p2k1*p2k2*pp1*qp1 + 2.*p1k1*p2k2*pp2*qp1 -
             2.*pk1*qk2*pp2*qp1 + 2.*pk1*pk2*p1p2*qp1 + 2.*p2k1*pk2*p1p2*qp1 -
             m2*p1k1*pk2*qp2 - m2*p2k1*pk2*qp2 + 4.*m2*pk1*p1k2*qp2 +
             m2*p1k1*p1k2*qp2 + m2*p2k1*p1k2*qp2 - 4.*m2*pk1*qk2*qp2 -
             2.*m2*p1k1*qk2*qp2 - 2.*m2*p2k1*qk2*qp2 + 2.*pk1*p1k2*pp1*qp2 +
             2.*p2k1*p1k2*pp1*qp2 - 2.*pk1*qk2*pp1*qp2 - 2.*p2k1*qk2*pp1*qp2 -
             2.*p1k1*p1k2*pp2*qp2 + 2.*p1k1*qk2*pp2*qp2 +
             k1k2*(-2.*m2*qp*(m2 + p1p2) - (m2*(m2 - pp1 + pp2) +
             (m2 + 2.*pp2)*p1p2)*qp1 + m2*(m2 + pp1 + pp2 + p1p2)*qp2))/2.0;

tr24 = (qp1*(-(m2*p2k1*pk2) - u2*p2k1*pk2 + m2*qk1*pk2 + m2*pk1*p1k2 +
	          m2*p2k1*p1k2 - m2*pk1*p2k2 - u2*pk1*p2k2 + m2*qk1*p2k2 - m2*pk1*qk2 -
	          m2*p2k1*qk2 + 2.*p2k1*p1k2*pp1 - 2.*p2k1*qk2*pp1 + 2.*qk1*p1k2*pp2 -
	          2.*qk1*qk2*pp2 - p1k1*(m2*pk2 + m2*p2k2 + 2.*(p1k2 - qk2)*pp2) -
	          2.*p2k1*p1k2*qp + 2.*p2k1*qk2*qp + 2.*pk1*p1k2*p1p2 - 2.*pk1*qk2*p1p2 +
	          2.*p2k1*pk2*qp1 + 2.*pk1*p2k2*qp1 + k1k2*(m2*pp1 +
	          pp2*(m2 + u2 - 2.*qp1) + m2*(m2 - qp + p1p2 - qp2)) -
	          2.*pk1*p1k2*qp2 + 2.*pk1*qk2*qp2))/2. +

	          u2*((m2*pk1*p1k2 - 2.*m2*pk1*p2k2 + m2*k1k2*pp1 + 2.*m2*k1k2*pp2 -
	          p1k1*(m2*pk2 + m2*p2k2 + 2.*(2.*p1k2 - qk2)*pp2) + m2*k1k2*p1p2 +
	          4.*pk1*p1k2*p1p2 - 2.*pk1*qk2*p1p2 + p2k1*(-2.*qk2*pp1 + p1k2*(m2 +
	          4.*pp1) - 2.*pk2*(m2 - qp1)) + 2.*pk1*p2k2*qp1 - 2.*k1k2*pp2*qp1)/4.) +

              m2*((2.*m2*qk1*pk2 - u2*qk1*pk2 - 2.*u2*pk1*p1k2 + 4.*m2*qk1*p1k2 -
              2.*u2*qk1*p1k2 - 2.*u2*pk1*p2k2 + 2.*m2*qk1*p2k2 - u2*qk1*p2k2 -
              2.*m2*pk1*qk2 + u2*pk1*qk2 - 2.*m2*p1k1*qk2 - 4.*m2*qk1*qk2 +
              2.*qk1*p1k2*pp1 + 2.*qk1*p2k2*pp1 - 4.*qk1*qk2*pp1 + 2.*p1k1*qk2*pp2 -
              4.*qk1*qk2*pp2 - 2.*p1k1*p1k2*qp + 2.*qk1*p1k2*qp - 2.*p1k1*p2k2*qp +
              2.*qk1*p2k2*qp + 2.*p1k1*qk2*qp + 2.*qk1*pk2*p1p2 + 2.*qk1*p1k2*p1p2 -
              2.*pk1*qk2*p1p2 - 4.*qk1*qk2*p1p2 + p2k1*(qk2*(-2.*m2 + u2 - 2.*pp1 + 2.*qp) -
              2.*pk2*(u2 - qp1) - 2.*p1k2*(u2 - qp1)) + 2.*pk1*p1k2*qp1 + 2.*pk1*p2k2*qp1 +
              4.*qk1*qk2*qp1 - 2.*p1k1*pk2*qp2 + 2.*qk1*pk2*qp2 - 2.*p1k1*p1k2*qp2 +
              2.*qk1*p1k2*qp2 + 2.*pk1*qk2*qp2 + 2.*p1k1*qk2*qp2 + k1k2*(-2.*m2*u2 +
              2.*pp2*(u2 - qp1) + 2.*m2*qp1 + qp*(2.*m2 + u2 + 2.*p1p2 - 2.*qp1 - 4.*qp2) +
              2.*m2*qp2 + u2*qp2 + 2.*pp1*qp2 - 2.*qp1*qp2))/4.) +

              u2*m2*((2.*p2k1*pk2 + qk1*pk2 + 3.*pk1*p1k2 + 3.*p2k1*p1k2 + 2.*qk1*p1k2 +
              2.*pk1*p2k2 + qk1*p2k2 - 3.*pk1*qk2 - 3.*p2k1*qk2 - p1k1*(pk2 + p2k2 +
              2.*qk2) + k1k2*(6.*m2 + 3.*pp1 - qp + 3.*p1p2 - qp2))/4.);

tr33 = -(qk2*(p1k1*(m4 + m2*pp1 - m2*pp2 + pp22 - (2.*m2 + pp1 + pp2)*p1p2) +
              p2k1*((m2 + pp1)*(2.*m2 - pp2) - (m2 + pp1 + 2.*pp2)*p1p2) +
              pk1*(m2*(m2 + pp1) - (2.*m2 + pp1)*pp2 - (m2 + pp2)*p1p2 + p1p22)));

tr34 = m2*pk1*p2k2*qp - m2*p2k1*p2k2*qp - p1k1*p2k2*pp1*qp -
             p2k1*p2k2*pp1*qp + 2.*p1k1*p1k2*pp2*qp + p2k1*p1k2*pp2*qp -
             p1k1*qk2*pp2*qp - 2.*pk1*p1k2*qp*p1p2 - p2k1*p1k2*qp*p1p2 +
             pk1*qk2*qp*p1p2 + 2.*p2k1*qk2*qp*p1p2 +
             qk1*(-2.*m2*p2k2*(m2 + pp1) + m2*pk2*pp2 + m2*qk2*pp2 +
             qk2*pp1*pp2 + m2*qk2*p1p2 - pk2*pp1*p1p2 + qk2*pp1*p1p2 +
             p1k2*(-(pp1*pp2) + m2*p1p2)) + m2*p1k1*p2k2*qp1 -
             m2*p2k1*p2k2*qp1 - pk1*p2k2*pp1*qp1 - p2k1*p2k2*pp1*qp1 -
             2.*p1k1*pk2*pp2*qp1 - p2k1*pk2*pp2*qp1 + p1k1*qk2*pp2*qp1 +
             2.*p2k1*qk2*pp2*qp1 + 2.*pk1*pk2*p1p2*qp1 + p2k1*pk2*p1p2*qp1 -
             pk1*qk2*p1p2*qp1 - m2*pk1*pk2*qp2 + m2*p2k1*pk2*qp2 -
             m2*p1k1*p1k2*qp2 + m2*p2k1*p1k2*qp2 - m2*pk1*qk2*qp2 -
             m2*p1k1*qk2*qp2 - 2.*m2*p2k1*qk2*qp2 + p1k1*pk2*pp1*qp2 +
             p2k1*pk2*pp1*qp2 + pk1*p1k2*pp1*qp2 + p2k1*p1k2*pp1*qp2 -
             pk1*qk2*pp1*qp2 - p1k1*qk2*pp1*qp2 - 2.*p2k1*qk2*pp1*qp2 +
             k1k2*(p1p2*(pp1*qp - m2*qp1) + pp2*(-(m2*qp) + pp1*qp1) +
             2.*m2*(m2 + pp1)*qp2);

tr44 = -(p2k1*(-(pk2*(pp1*(u2 + qp) + m2*(u2 + 2.*qp) +
             (m2 - qp)*qp1 + qp12)) - p1k2*(m2*u2 + m2*qp + qps +
             (2.*m2 - qp)*qp1 + pp1*(u2 + qp1)) + qk2*(m2*(m2 + u2 + qp) +
             (m2 - 2.*qp)*qp1 + pp1*(m2 + u2 + qp + qp1))));

matr2e = C1*C1*D1*D1*tr11 - C1*C1*D1*D2*tr13 + C1*C1*D2*D2*tr33;
	  
matr2mu =C2*C2*D1*D1*tr22 - C2*C3*D1*D2*tr24 + C3*C3*D2*D2*tr44;
	  
matr2emu = C1*C2*D1*D1*tr12 - C1*C3*D1*D2*tr14 - C1*C2*D1*D2*tr23 + C1*C3*D2*D2*tr34;
	  
matr2 = matr2e + matr2mu + matr2emu;
 
In this notations the normalized spin-averaged amplitude
of the process squared $\overline{|\tilde{F}|^2} \equiv matr2$. 

An uncertainty of the branching ratio for n trials is calculated by using a formula

\begin{equation}
\sigma = \big (\frac {\overline {W^2} - (\overline W)^2}{n} \big )^{1/2} 
\end{equation}
where average weights are given by
$\overline W = \sum_{i=1}^n W_i/n$ and $\overline {W^2} = \sum_{i=1}^n W_i^2/n$.
\end{document}